\documentclass[12pt]{article}
\usepackage[utf8]{inputenc}
\usepackage{mathpazo}
\usepackage{fullpage}
\usepackage{graphicx}
\usepackage[sort&compress,numbers]{natbib}
\usepackage{lipsum}
\usepackage{hyperref}
\usepackage{enumitem,amssymb}
\usepackage[total={6.5in,9in},top=1in,headsep=0.1in,headheight=1in]{geometry}
\usepackage[usenames,dvipsnames]{xcolor}
\usepackage[vietnamese,english]{babel}
\usepackage{aas_macros}

\graphicspath{{figures/}}

\DeclareUnicodeCharacter{2009}{\,}

\newcommand\snowmass{
\begin{center}
  \rule[-0.2in]{\hsize}{0.01in}\\
  \rule{\hsize}{0.01in}\\
  \vskip 0.1in
  Submitted to the Proceedings of the US Community Study\\ 
  on the Future of Particle Physics (Snowmass 2021)\\
  \rule{\hsize}{0.01in}\\
  \rule[+0.2in]{\hsize}{0.01in}\\[-2em]
\end{center}
}

\usepackage[firstpage=true]{background}
\backgroundsetup{contents={\parbox{6.5in}{\snowmass}}, scale=1,placement=top,opacity=1,color=black,position={3.25in,1.2in}}

\usepackage{fancyhdr}
\fancypagestyle{plain}{%
  \fancyhf{}%
  \fancyhead[C]{}
  \fancyfoot[C]{\thepage}
}

\fancypagestyle{empty}{%
  \fancyhf{}%
  \fancyhead[C]{{\it Snowmass2021 Cosmic Frontier: CMB Measurements White Paper}}
  \fancyfoot[C]{\thepage}
}
\title{Snowmass2021 Cosmic Frontier: Cosmic Microwave Background Measurements White Paper}
\date{}

\usepackage{authblk}

\renewcommand\Affilfont{\scriptsize}
\makeatletter
\renewcommand\AB@affilsepx{, \protect\Affilfont}
\makeatother

\author[1,2]{Clarence L.~Chang}
\author[3]{Kevin M.~Huffenberger}
\author[4,1]{Bradford A. ~Benson}
\author[5,6]{Federico Bianchini}
\author[7]{Jens Chluba}
\author[8]{Jacques Delabrouille}
\author[9]{Raphael Flauger}
\author[10]{Shaul Hanany}
\author[11]{William C.~Jones}
\author[12]{Alan J.~Kogut}
% ...

\affil[1]{Department of Astronomy and Astrophysics, University of Chicago}
\affil[2]{Argonne National Laboratory, Lemont, IL 60439, USA}
\affil[3]{Department of Physics, Florida State University}
\affil[4]{Fermi National Accelerator Laboratory}
\affil[5]{Kavli Institute for Particle Astrophysics and Cosmology, Stanford University}
\affil[6]{SLAC National Accelerator Laboratory, Menlo Park, CA 94025, USA}
\affil[7]{Jodrell Bank Centre for Astrophysics at The University of Manchester}
\affil[8]{Centre Pierre Bin\'etruy International Research Laboratory, CNRS, UC Berkeley and LBNL, Berkeley, CA 94720, USA}
\affil[9]{UC San Diego, Department of Physics, La Jolla, CA, 92093}
\affil[10]{School of Physics and Astronomy, University of Minnesota}
\affil[11]{Department of Physics, Princeton University}
\affil[12]{Goddard Space Flight Center}
% ...

\begin{document}

\maketitle

% Pages for endorsements and comments
%\href{https://docs.google.com/spreadsheets/d/1YOu9aToem6cUg2X-fQWlUz0ll5_O4nSiw9cIq0g7pwU/edit?usp=sharing}{Add your name to the endorsers list} or \href{https://docs.google.com/document/d/1icLyPyACnYvtEwpVun3TTQovK9WLBjNRAmagwwiF2CM/edit?usp=sharing}{leave a comment on the draft}.

\section{Introduction}
%This whitepaper addresses the status and opportunities for Cosmic Frontier science from measurements of the Cosmic Microwave Background (CMB).  Recently, in \textit{Pathways to Discovery in Astronomy and Astrophysics for the 2020s,} the decadal survey report strongly endorsed CMB science in general and the multitude of discoveries it can offer \cite{2021pdaa.book......}, and recommended the CMB-S4 project as one of the top priorities for ground-based facilities.  
%In this and other whitepapers \citep[e.g.][]{Snowmass2021:Inflation,Snowmass2021:LightRelics,Snowmass2021:CosmoLabNeutrinos, Snowmass2021:DarkMatter}, we discuss the substantial scientific achievements that we expect from future CMB experiments, and we advocate support both for CMB projects in general and for the researchers who are commencing design, development, and analysis programs for them.  

%In this whitepaper, after describing the science case (section \ref{sec:science}), we survey the landscape of current and near-term CMB measurements (section \ref{sec:current_and_upcoming}).  We highlight the CMB-S4 project, a community-wide effort to improve our CMB capabilities by an order of magnitude over the previous generation (section \ref{sec:cmb-s4}).  We also discuss the CMB-HD concept, an effort to enable measurements of yet higher resolution and sensitivity (section \ref{sec:cmb-hd}).  For context, we also discuss the complementary capabilities of future space-based and balloon observations (section \ref{sec:space}).

The CMB is foundational to our understanding of modern physics and continues to be a powerful tool driving our understanding of cosmology and particle physics. In this and other whitepapers \citep[e.g.][]{Snowmass2021:Inflation,Snowmass2021:LightRelics,Snowmass2021:CosmoLabNeutrinos, Snowmass2021:DarkMatter}, we outline the broad and unique impact of CMB science for the High Energy Cosmic Frontier in the upcoming decade ($\sim$2025--2035). We also describe the progression of ground-based CMB experiments, which shows that the community is prepared to develop the key capabilities and facilities needed to achieve these transformative CMB measurements. Recently, in \textit{Pathways to Discovery in Astronomy and Astrophysics for the 2020s,} the decadal survey report strongly endorsed CMB science in general and the multitude of discoveries it can offer \cite{2021pdaa.book......}, and recommended the CMB-S4 project as one of the top priorities for ground-based facilities. This recommendation reflects both the tremendous impact of CMB science, and the readiness of the CMB community to carry out this program.

In this paper, after describing the potential of CMB science (section \ref{sec:science}), we present an overview of current and near-term CMB measurements (section \ref{sec:current_and_upcoming}). We highlight the CMB-S4 project, a community-wide effort to extend our CMB capabilities by an order of magnitude over the previous generation (section \ref{sec:cmb-s4}) and thus target key scientific thresholds in the upcoming decade. We also discuss opportunities for CMB science in the following decade including  the emerging CMB-HD concept (section \ref{sec:cmb-hd}).  For context, we discuss complementary capabilities of future space-based and balloon observations (section \ref{sec:space}).

\section{CMB Science on the Cosmic Frontier}
\label{sec:science}

\begin{figure}[t]
    \centering
    \includegraphics[width=0.95\textwidth]{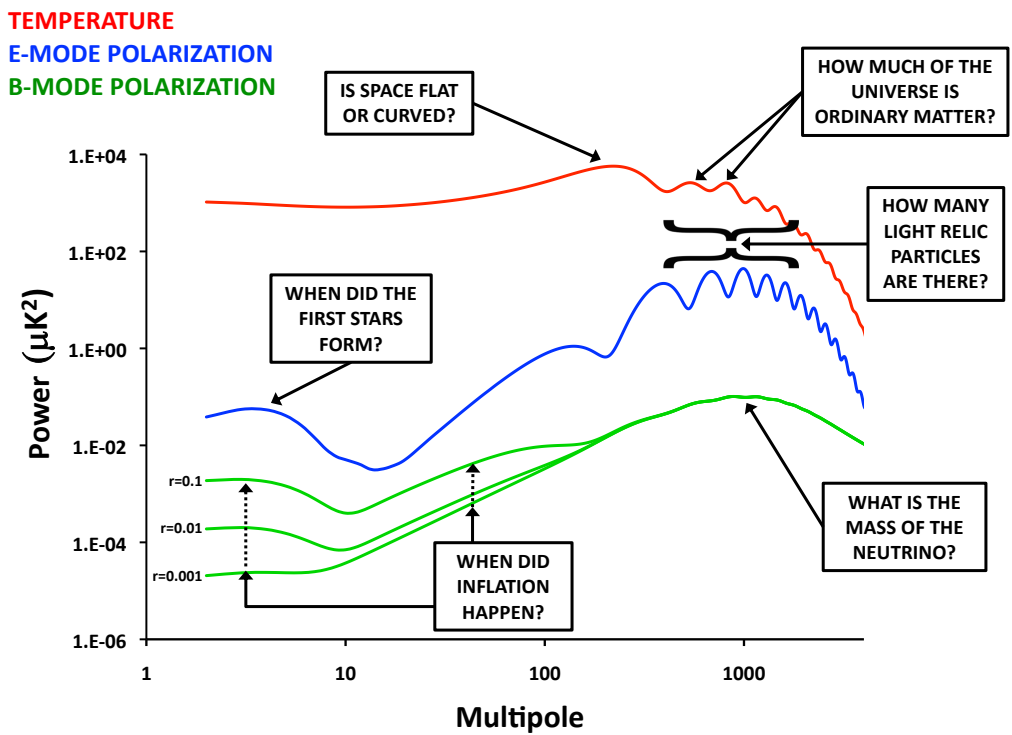}
    \vskip -36pt
    \caption{Some scientific cases that are revealed by different portions of the CMB temperature and $E$/$B$-polarization angular power spectra. Improved measurements connect to new constraints on high-energy physics and astrophysics, as approximately shown.  The multipole axis indicates the angular scale of the anisotropies on the sphere, left-to-right, from hemisphere-scale to arcminute-scale fluctuations.}
    \label{fig:cmb_power_spectra}
    \vskip -14pt
\end{figure}

%The CMB radiation yields insights into fundamental physics through two main mechanisms.  In the afterglow of the Big Bang, the ``primary'' CMB temperature and polarization anisotropies record a snapshot of the conditions at the time that the Universe first becomes transparent.  Those conditions depend on the Universe's initial conditions, its contents, and its evolution to that point.  In the ``secondary'' anisotropies, we recognize electron scattering interactions and gravitational lensing as the CMB illuminates all the later-developing structures in the Universe. Individually and in concert, these mechanisms provide tools for us to constrain fundamental physics related to inflation, neutrinos, light relic particles, dark matter, dark energy, and other elements of beyond-the-standard-model physics.
The CMB radiation yields insights into fundamental physics through two main mechanisms. First, the ``primary'' CMB temperature and polarization anisotropies record a snapshot of the conditions at the time that the Universe first becomes transparent. At degree angular scales and larger (see Fig.~\ref{fig:cmb_power_spectra}), the patterns in the CMB are generated by physics that preceded the hot radiation-dominated era. These primordial features directly probe the initial conditions of the Universe. At smaller scales (see Fig.~\ref{fig:cmb_power_spectra}), the CMB exhibits patterns arising from acoustic oscillations, an exquisite tracer of the thermal evolution of the early universe. Precision measurements of the CMB at these angular scales translate into precision measurements of the energy composition of the early universe. The second mechanism is the “secondary” anisotropies, where electron scattering interactions and gravitational lensing reveal the later-developing structures in the Universe. Individually and in concert, these mechanisms provide tools for us to constrain fundamental physics related to inflation, neutrinos, light relic particles, dark matter, dark energy, and other elements of beyond-the-standard-model physics.

\begin{figure}[t]
\begin{center}
\includegraphics[width=\textwidth]{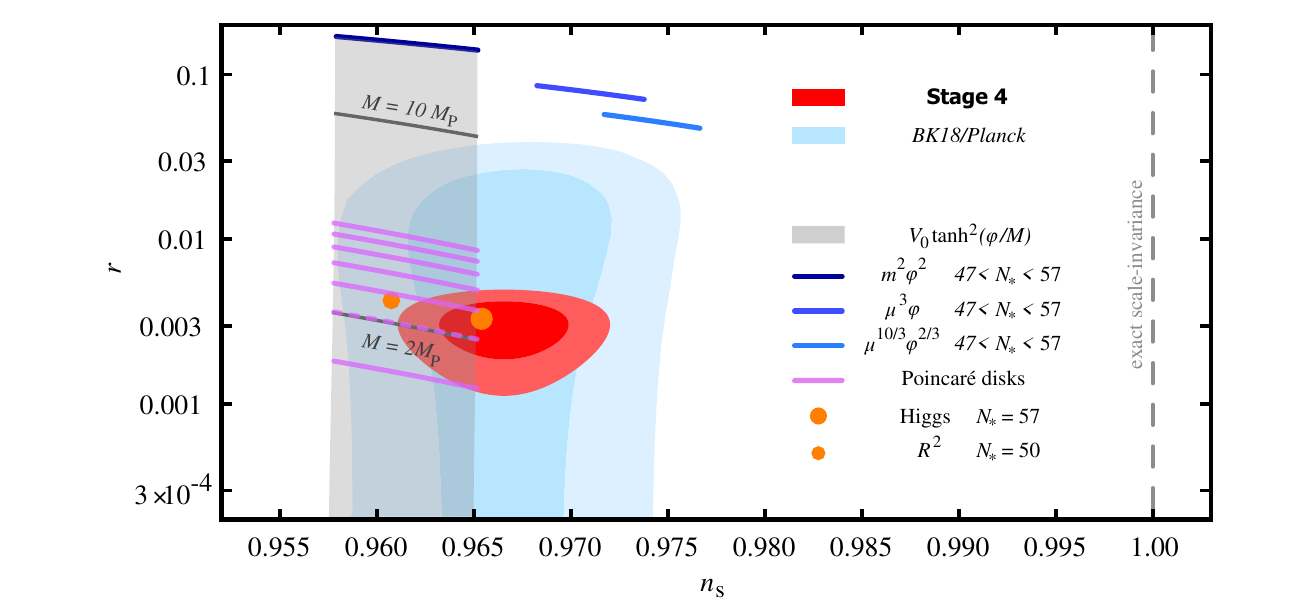}
\end{center}
\caption[]{Model space of the $n_{\rm s}$--$r$ plane for inflation.   
 Shown are the current best constraints from a combination of the BICEP2/{\em Keck Array\/} experiments and Planck~\cite{2021PhRvL.127o1301A}, and the constraining power of a Stage 4 experiment for a model with $r=0.003$. Models that naturally explain the observed departure from scale invariance separate into two viable classes: monomial and plateau. The monomial models ($V(\phi)=\mu^{4-p}\phi^{\,p}$) are shown for three values of $p$ as blue lines for $47<N_\ast<57$ (with the spread in $N_\ast$ reflecting uncertainties in reheating, and smaller $N_\ast$ predicting lower values of $n_{\rm s}$). The simplest realization of this class is now disfavored. The plateau models include the $\tanh^2$ form (gray band) as an example, as this form arises in  a sub-class of $\alpha$-attractor models~\cite{Kallosh:2013hoa}. Some particular realizations of physical models in the plateau class are also shown: the Starobinsky model~\cite{Starobinsky:1980te} and Higgs inflation~\cite{Bezrukov:2007ep} (small and large orange filled circles, respectively) and Poincar\'e disks. The differing choices of $N_\ast$ for Higgs and Starobinsky reflect differing expectations for reheating efficiency. 
}
\label{fig:nsrp01}
\end{figure}

\paragraph{Inflation.}  %Cosmic inflation is a successful paradigm for describing the phase of the very early universe that precedes the hot radiation-dominated era, but we so far lack definitive evidence in favor of this inflationary epoch.  Observation of primordial gravitational waves produced during the rapid inflationary expansion could provide such evidence. 
Cosmic inflation posits an epoch of accelerated cosmic expansion during the very early universe that precedes the hot radiation-dominated era~\cite{Snowmass2021:Inflation}. This relatively simple paradigm successfully describes all of our current observations of the primordial universe and predicts unique signals to further test and constrain the theory. A key prediction of cosmic inflation is the existence of a background of primordial gravitational waves. These tensor waves imprint a distinct pattern in the CMB polarization. Precision measurement of the CMB $B$-mode polarization at degree angular scales is the most powerful experimental technique for searching for this signature. Observation of this signal would be a watershed detection providing definitive evidence in favor of this inflationary epoch. Importantly, the strength of the signal, parameterized as $r$ (the ratio of power in tensor to scalar perturbations), depends on the underlying physics of cosmic inflation. 
Many models of inflation that naturally explain the deviation from scale invariance of the scalar perturbations ($n_s < 1$), while having a characteristic scale larger than the Planck mass, have tensor-to-scalar ratios $r>0.001$, and well-motivated subclasses have $r>0.003$. (See Fig.~\ref{fig:nsrp01}.)  %These tensor gravitational waves would produce $B$-mode polarization 
Measuring CMB B-mode polarization to this level is a key objective of a next-generation CMB facility \cite{Abazajian:2019eic,2022ApJ...926...54A} and will have profound implications for our understanding of physics at the highest energies and earliest moments. 

While the tightest constraints are expected from observations of $B$-mode polarization, the inflationary paradigm gives rise to other observables that provide avenues for studying inflationary physics. %a complementary probe of primordial gravitational waves is possible via 
One possible signal uses high-resolution polarized Sunyaev-Zel'dovich observations in cross-correlation with galaxy surveys as an indirect probe of low-$\ell$ CMB polarization \cite{CMB-HD-Snowmass}. Another signal is primordial non-Gaussianities, where the CMB provides a clean, robust observable.
%Furthermore, the CMB is the cleanest and most robust observable for studies of primordial non-Gaussianities that may be produced during or after a period of inflation.  Non-Gaussianities also arise in models with $B$-modes below the limit of detectability.  
Alone and in combination with galaxy-survey measurements from Rubin-LSST, DESI, and future galaxy imaging and redshift surveys, non-detection of non-Gaussianity would be a tight constraint, potentially ruling out broad swaths of the inflationary model space.  On the other hand, a detection could demonstrate what type of models are valid.
Isocurvature fluctuations for cold dark matter, baryons, or neutrinos must contribute less than one percent of the total power \cite{Enqvist:2000hp,MacTavish:2005yk,Dunkley:2008ie,Planck:2013jfk,Ade:2015lrj,Akrami:2018odb}, but the discovery of small isocurvature fluctuations by future CMB measurements would, for example, probe inflation under the curvaton, axion isocurvature, or compensated isocurvature perturbation scenarios.    
Precision CMB polarization measurements significantly enhance our ability to detect non-trivial features in the primordial power spectrum.  Such features would be signs of the physics of the inflationary epoch \cite{Slosar:2019gvt,2003MNRAS.342L..72B, 2011JCAP...08..031G, 2012ApJ...749...90H, 2013PhRvD..87h3526A, 2014JCAP...01..025H, 2014PhRvD..89j3502D, 2016PhRvD..93b3504M, 2016JCAP...09..009H,Dvorkin:2009ne,Dvorkin:2010dn,Dvorkin:2011ui,Obied:2017tpd,Obied:2018qdr}.

\paragraph{Light relics and neutrino physics.}
%High-resolution and wide-area CMB polarization measurements will provide a precise measurement of the light relic density and allow us to probe the thermal history of the Universe in exquisite detail.  This measurement will potentially reveal new light relic particles.  Such particles are ubiquitous in well-motivated extensions to the standard model, while a measurement consistent with the standard model would place broad constraints on the physics of the dark sector~\cite{Green:2019glg,Snowmass2021:LightRelics}. CMB observations thereby serve as a valuable complement to laboratory and astrophysical probes of dark sector physics. The sensitivity of a CMB survey to light relics can be understood via the parameter $N_{\rm eff}$, the effective number of light species. Sensitivity to $N_{\rm eff}$ at a level of $\Delta N_{\rm eff}\sim0.03$ would constrain thermal light relics that have freeze-out temperatures up to the QCD phase transition.
%The increase in sensitivity to the light relic density offered by upcoming CMB surveys allows us to search for new light relic particles with extremely weak couplings to the particles of the standard model.  
High-resolution and wide-area CMB polarization measurements map the thermal history of the early universe in exquisite detail. From these measurements, the CMB can determine, with precision, the composition of our early universe including constraining the energy density in dark radiation. The dark radiation energy density is typically paramaterized as $N_{\rm eff}$, the effective number of light species. In the standard model, neutrinos are the only contribution to $N_{\rm eff}$ with a calculated value of $N_{\rm eff}^\mathrm{SM}$=3.045. A robust measurement deviating from this predicted value is evidence of new physics. Examples of non-standard model contributions to $N_{\rm eff}$ include a stochastic background of gravitational waves~\cite{Boyle:2007zx, Stewart:2007fu, Meerburg:2015zua} generated by especially violent sources~\cite{Caprini:2018mtu, Adshead:2018doq, Amin:2019qrx} or thermal relics from new light degrees of freedom that are frequently predicted in various well-motivated extensions to the standard model. With regards to the latter, constraints on $N_{\rm eff}$ have broad implications on the physics of the dark sector~\cite{Green:2019glg,Snowmass2021:LightRelics} complementing laboratory and astrophysical probes of dark sector physics. Sensitivity to $N_{\rm eff}$ at a level of $\sigma(N_{\rm eff})\sim0.03$ would constrain thermal light relics that have freeze-out temperatures up to the beginning of the QCD phase transition. Sensitivity to $N_{\rm eff}$ at a level of $\sigma( N_{\rm eff})\sim0.01$ would constrain all thermal light relics.

\begin{figure}
\centering
\includegraphics[width=0.9\textwidth, trim=0in 2.6in 0in 2.6in]{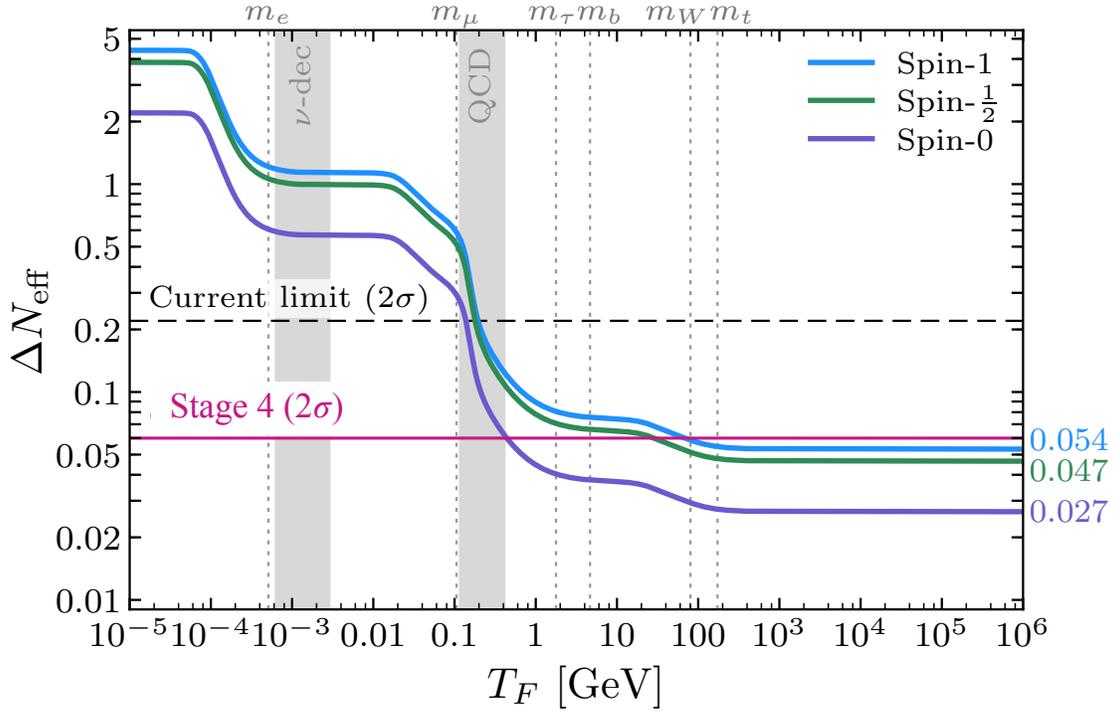}
%\vspace{0.1cm}
\caption{Contributions of a single massless particle, which decoupled from the Standard Model at temperature $T_F$, to the effective number of relativistic species, $N_{\rm eff} = N_\mathrm{eff}^\mathrm{SM} + \Delta N_{\rm eff}$, with the SM expectation $N_\mathrm{eff}^\mathrm{SM} = 3.045$ from neutrinos. The dashed line shows the 2$\sigma$ limit from a combination of current CMB, BAO, and Big Bang nucleosynthesis (BBN) observations~\cite{Aghanim:2018eyx}.  The purple line shows the projected sensitivity of a ``Stage 4'' CMB instrument and illustrating the power to constrain light thermal relics. The displayed values on the right are the observational thresholds for particles with different spins and arbitrarily large decoupling temperatures.}%\vspace{-9pt}
\label{fig:deltaNeff}
\end{figure}

%The same type of CMB data that allows for a precise measurement of the light relic density also enables cosmological probes of the sum of neutrino masses~\cite{Snowmass2021:CosmoLabNeutrinos}.  
CMB data also enables cosmological probes of the sum of neutrino masses~\cite{Snowmass2021:CosmoLabNeutrinos}. Massive neutrinos were relativistic at early times but contribute to the non-relativistic matter density today.  Their large thermal velocities suppress the clustering of matter on scales smaller than their free streaming length, compared to a model containing massless neutrinos.  CMB observations allow us to search for this suppression using measurements of weak gravitational lensing and the evolution of the number density of galaxy clusters. These measurements are complementary to other probes of cosmological structure growth, and combining CMB surveys with galaxy surveys enables cross-correlations that enhance the constraining power of both. Sensitivity to the summed neutrino mass at the level of $\sigma(\Sigma {\rm m}_\nu)<0.02$~eV would constitute a cosmological measurement at the level of the largest neutrino mass splittings, complementing laboratory measurements and contributing to the understanding of neutrino masses and hierarchy. %Cosmological measurements of neutrino mass inform laboratory constraints on the masses and mass hierarchy.

\paragraph{Dark Matter.}
CMB measurements have firmly established that a significant fraction of the energy budget of the Universe is in the form of non-baryonic dark matter.  The next generation of CMB measurements provides the chance to probe the particle properties of dark matter.   

Dark matter--baryon scattering interactions, if present, would impart a subtle drag force that dampens acoustic oscillations in the CMB, with the largest effects on small scales.  Such a search is promising for low-mass (sub-GeV) WIMP-like particles because the lower mass implies a larger number of scatterers, in contrast to the GeV and higher scales accessible by traditional nuclear recoil experiments in the lab.  In this regime, the CMB is sensitive to cross-sections above the nuclear scale.  

Dark matter that interacts with any type of relativistic dark radiation at early times suppresses growth due to the radiation pressure, and puts a signature into the CMB acoustic oscillations.  For dark matter candidates based on QCD axions or axion-like particles, the CMB also provides tools for detection and probing the physics.  In the ultra-light axion scenario, CMB power spectra measurements can constrain the density, mass scale, and axion decay constant \citep[e.g.][]{2017PhRvD..95l3511H}.

CMB experiments can also constrain or discover axion-like particles by observing the resonant conversion of CMB photons into axion-like particles in the magnetic fields of galaxy clusters. Nearly massless pseudoscalar bosons, often generically called axions-like particles, appear in many extensions of the standard model~\cite{PhysRevLett.38.1440,Weinberg:1977ma,PhysRevLett.40.279,Svrcek:2006yi,Arvanitaki:2009fg,Acharya:2010zx}. A detection would have major implications both for particle physics and for cosmology, not least because axion-like particles are a well-motivated dark matter candidate.

Additionally, CMB measurements can constrain or discover axion-like particles by measuring the time-dependent CMB polarization rotation. Ultralight axion-like dark-matter fields that couple to photons via $g_{a\gamma}$ cause a time-dependent photon birefringence effect which manifests as a temporal oscillation of the local CMB polarization angle (i.e., a local $Q\leftrightarrow U$ oscillation in time)~\cite{Fedderke:2019ajk}.
This rotation effect is in-phase across the sky, and the oscillation period is fixed by the mass of the axion-like particle (a fundamental physics parameter) to be at observable timescales of $\sim$months to $\sim$hours for masses in the range $10^{-21}\,\textrm{eV}\lesssim m_a \lesssim 10^{-18}\,\textrm{eV}$. Since the effect is a time-dependent oscillation of the \emph{observed} CMB polarization pattern, searches for this effect are not limited by cosmic variance.

CMB lensing can enable the measurement of the small-scale matter power spectrum from weak gravitational lensing using the primordial CMB as a backlight. This measurement is a direct probe of the dark matter distribution, free of the use of baryonic tracers. Lensing will greatly limit the allowed model-space for dark matter~\cite{Hlozek:2016lzm,Hlozek:2017zzf,Nguyen:2017zqu}, constraining ultra-light axions, warm dark matter, self-interacting dark matter, and any other dark matter model that alters the matter power spectrum on small scales~\cite{Markovic:2013iza,Hlozek:2016lzm,Hlozek:2017zzf,Li:2018zdm,Nguyen:2017zqu}.

\paragraph{Dark Energy.}
Dark energy becomes prominent in the late Universe, much later than the imprinting of the primordial CMB fluctuations, but its effect on the expansion history sets the angular scale for the fluctuations we observe.  Observables of the secondary anisotropies are sensitive to the impact of dark energy on observables of the growth of structure, chiefly the weak lensing field, counts of Sunyaev-Zel'dovich galaxy clusters, and the kinematic-Sunyaev-Zel'dovich velocity field.  An early-time component or type of dark energy could alter the expansion rate before recombination, decreasing the physical sound horizon, and thus decreasing our inferred distance to the CMB and increasing our inferred Hubble constant, an effect that could potentially resolve the discrepancy between CMB and local supernova $H_0$ measurements~\cite{Riess:2021jrx}.  The presence of an early dark energy component is constrained by a combination of CMB and large-scale structure data.

\paragraph{Other beyond-the-standard-model physics.} 
CMB experiments are also sensitive to potential parity-violating physics that could arise from fields not included in the standard model of particle physics.  A scalar field that couples to photons via a Chern-Simons term in the Lagrangian would rotate the linear polarization of the CMB light as it travels over cosmological distances.  That cosmic birefringence rotation converts some of the even parity $E$-type polarization pattern to an odd-parity $B$-type pattern, and induces $TB$ and $EB$ correlations that are not present without the parity violating mechanism.  These can be measured in CMB data with 2- and 4-point statistics.
In a similar way, primordial magnetic fields partially convert (by Faraday rotation) the $E$ polarization to $B$, and can be detected by similar means. The measurement of cosmic birefringence constrains very light axion-like particles of $m_a \lesssim 10^{-28}$\,eV \cite{Harari:1992:axion,Carroll:1998,Li:2008,Pospelov:2009,Capparelli:2019:CB}, the axion string network \cite{Agrawal:2019:biref}, axion dark matter \cite{Liu:2016dcg}, general Lorentz-violating physics in the context of Standard Model extensions \cite{Leon:2017}, and primordial magnetic fields (PMFs) through the Faraday rotation \cite{Kosowsky:1996,Harari:1997,Kosowsky:2004:FR}. 

These types of measurements can be used to bolster the science cases listed above.  For example, the measurement of scale-invariant inflationary magnetic fields via measurements of anisotropic birefringence~\cite{Mandal:2022tqu} provides an alternate path to probing inflation. If the detected amplitude is above $0.1\,\mathrm{nG}$ on Mpc scales, then the origin of magnetic fields observed in galaxies today must be from inflation.  This would be compelling evidence for inflation itself, since only an inflationary mechanism could generate such a strong, scale-invariant magnetic field on Mpc scales.

\section{Current and Upcoming Facilities}
\label{sec:current_and_upcoming}
\begin{figure}[t]
    \centering
    \includegraphics[width=0.8\textwidth]{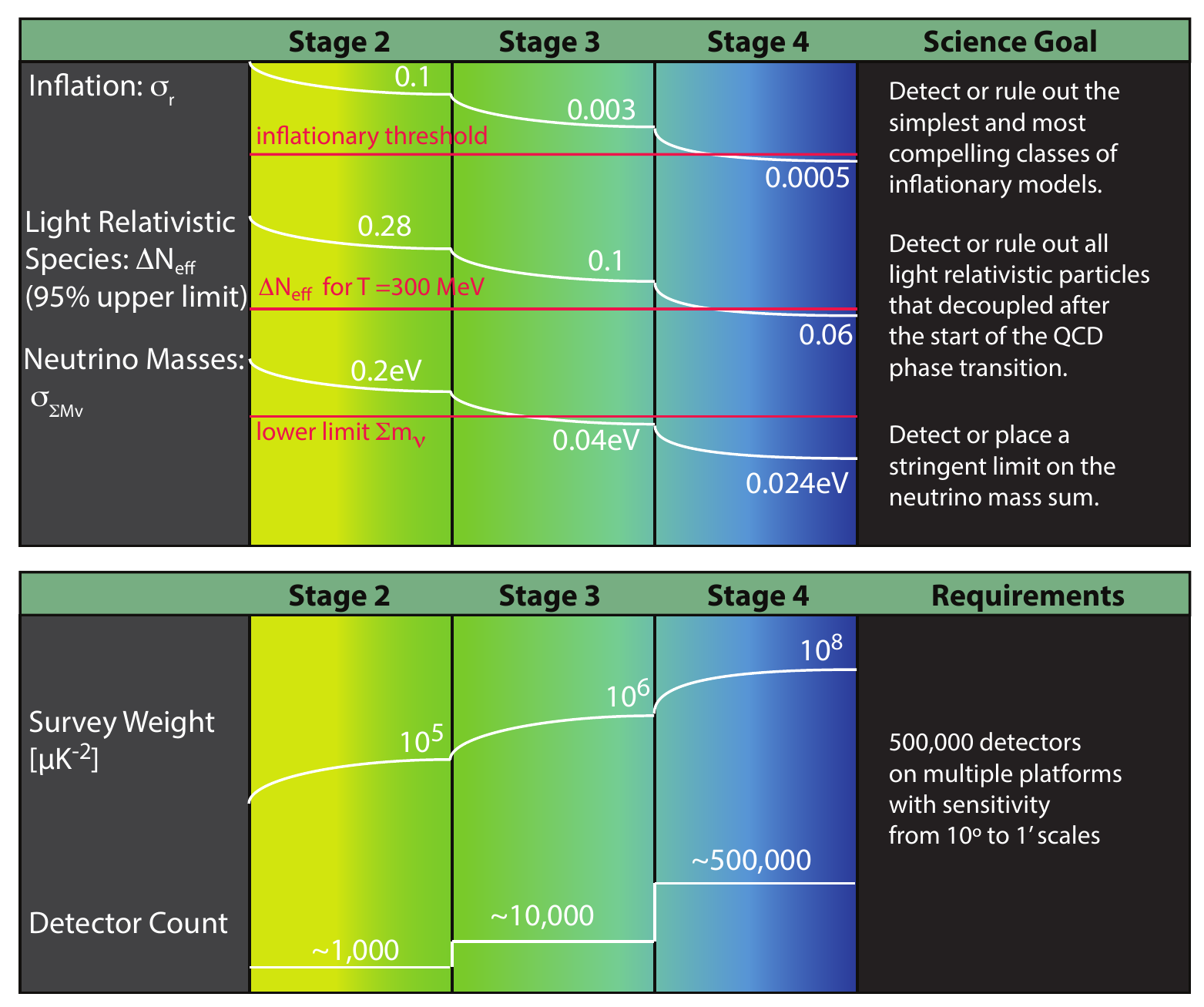}
    \caption{Illustration of the ``Stage'' classification of CMB experiments. Bottom: The size and sensitivity (i.e. survey weight) of a CMB experiment scales with the number of single-moded detectors. ``Stage 2'' experiments have $O$(1000) detectors, ``Stage 3'' experiments have $O$(10,000) detectors and ``Stage 4'' experiments have $O$(100,000) detectors. Top: The increased sensitivity of the larger experiments results in greater science reach with ``Stage 4'' experiments crossing several scientific thresholds.}
    \label{fig:CMB_stages}
\end{figure}

The CMB anisotropy and its polarization are measured through repeated observations of the sky using mm-wave telescopes.  Typical campaigns with ground-based instruments span multiple years. Because the CMB is a diffuse signal with features at large angular scales, CMB telescopes have been developed to include significant baffling and shielding providing tight control and mitigation of systematic effects. The angular resolution of a CMB measurement is fundamentally limited by diffraction and, for single-moded systems, scales with the size of the telescope aperture. Smaller resolution requires larger telescopes, and commensurate increase in the associated shielding and baffling. The mm-wave cameras for CMB experiments are sub-Kelvin radiometers equipped with cryogenic optics, internal baffling and superconducting detectors and readout. The detectors are typically single-moded bolometers with an optical bandwidth of $\sim$25\%. The dominant noise for an individual CMB detector is the shot noise of the absorbed photons which come from the cryostat, the sky signal, and, for terrestrial observations, the atmosphere. These photon fluctuations fundamentally limit additional improvements of individual detector sensitivity---increasing the sensitivity for a CMB instrument requires increasing the number of detectors. This connection between the size of the detector payload and the overall instrument sensitivity provides a general framework for categorizing CMB experiments. Over the past decade, the community has classified ground-based instruments into ``Stage 2'' ($O$(1000) detectors), ``Stage 3'' ($O$(10,000) detectors) and ``Stage 4'' ($O$(100,000) detectors) experiments. Figure~\ref{fig:CMB_stages} illustrates the connection between the size of the experiment and its scientific reach.

\begin{figure}
    \centering
    \includegraphics[width=5.8in]{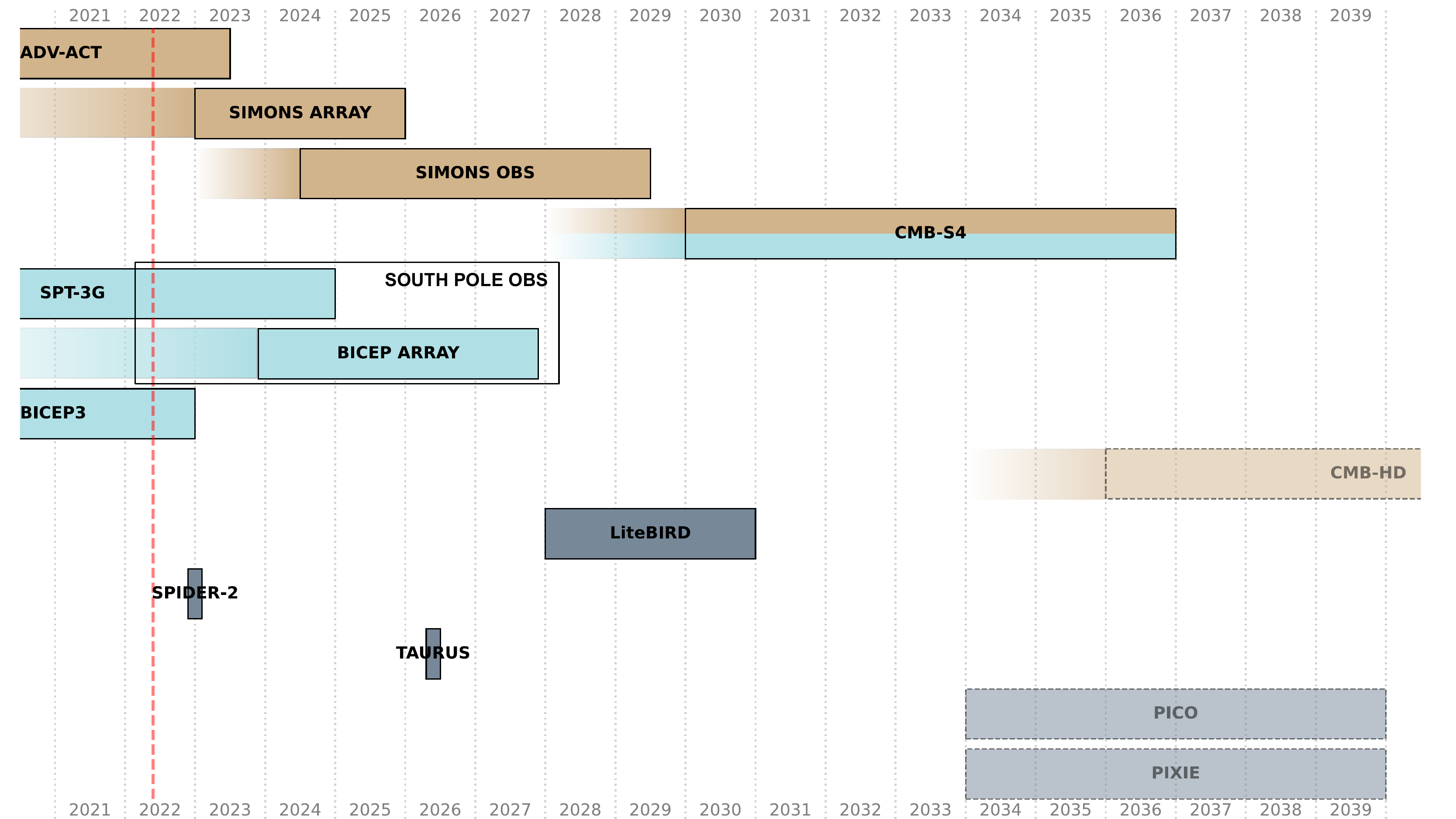}
    \caption{Timeline of current and future ground-based CMB experiments. For context, the timeline also includes a few sub-orbital and satellite experiments.}
    \label{fig:CMBexp_timeline}
\end{figure}

Currently, the field of ground-based CMB is in ``Stage 3'' and transitioning to ``Stage 4.'' In this section, we review the present-day landscape of CMB-experiments and then the facilities coming online in the next few years. Figure~\ref{fig:CMBexp_timeline} presents a current timeline for CMB experiments from 2020--2040 and shows, for ground-based instruments, the increasing experiment and collaboration size and the corresponding consolidation of the experimental landscape.

\subsection{Current Ground-based Measurements (``Stage 3'')}
%We define the current generation of ground-based experiments as experiments that are currently taking data and refer to them generally as third-generation or ``Stage 3'' experiments. Typically these experiments are one of two main types: small-aperture telescopes (searching for primordial $B$-modes) and large-aperture telescopes (measuring the small-scale CMB power spectrum, gravitational lensing, galaxy clusters, and other astrophysical objects). Spectral lines in the atmosphere from 30-300~GHz (e.g. from water vapor) contaminate CMB observations, so Stage~3 experiments have primarily operated out of one of two sites: the South Pole and the Atacama Desert in Chile, which are among the driest deserts on Earth.
The current generation of ground-based experiments (either taking data or under construction) are ``Stage 3'' experiments. To minimize the contribution from spectral lines in the atmosphere between 30-300~GHz (e.g. from water vapor), these experiments operate from one of two sites: the South Pole and the Atacama Desert in Chile, which are among the driest deserts on Earth.

%The current generation of ground-based CMB telescopes, those that are taking data now, are of two main types and are based at one of two sites.  We refer to them collectively as third-generation or ``stage 3.''  The two types of telescopes are small-aperture telescopes (searching for primordial $B$-modes with low systematics) and large-aperture telescopes (measuring the small-scale CMB power spectrum, gravitational lensing, and galaxy clusters and other astrophysical objects).  Spectral lines from water vapor contaminates these 30-300 GHz measurements, so the two sites, at the South Pole and in the Atacama Desert in Chile, are among the driest deserts on Earth.  

Experiments like the 10-m South Pole Telescope (SPT)~\cite{SPT-3G:2014dbx} and the 6-m Atacama Cosmology Telescope (ACT)~\cite{ACT:2020gnv} use large  aperture telescopes that provide arcminute angular resolution, enabling measurements of the small-scale CMB power spectrum, gravitational lensing, galaxy clusters and other astrophysical objects. These experiments have similar science goals, but recently have used survey strategies that take advantage of the natural conditions of each site.  The sky at the South Pole never sets, but instead rotates around the Earth's axis. SPT thus focuses on observing smaller, always-visible patches, to a  more-sensitive observing depth. ACT is at mid-latitude (23$^\circ$ S), so it can access large areas of the sky. ACT thus focuses on observing relatively larger areas of the sky to shallower depths. These optimized strategies enable SPT to measure the CMB temperature, polarization, and lensing modes to lower noise levels, and find smaller galaxy clusters and fainter point sources. Meanwhile, ACT can measure more CMB and lensing modes, reducing the cosmic variance on the measurements, find a larger number of large and bright but rare clusters and sources, and have significant overlap with a number of optical surveys.

% SATs
The 
BICEP/Keck~\cite{BicepKeck:2022dtc} series of experiments at the South Pole, the POLARBEAR/Simons Array~\cite{POLARBEAR:2022dxa} series of experiments in the Atacama, and the ABS/CLASS~\cite{Essinger-Hileman:2014pja} series of experiments in the Atacama, focus on larger angular scales to search for inflationary $B$-modes, but differ in important respects. The BICEP/KECK telescopes are refractors, while POLARBEAR/Simons Array and ABS/CLASS are reflectors.  Both BICEP/KECK and POLARBEAR/Simons Array search in the multipole range of the recombination bump ($\ell \sim 80$) in the $B$-mode power spectrum as did ABS  during its operation (2012-2014). CLASS searches at larger angular scales ($\ell < 10$) for the reionization bump in the  $B$-mode power spectrum.  CLASS also seeks to improve constraints on the reionization optical depth.

% LATs
%Among the large-aperture telescopes, the 10-m South Pole Telescope (SPT)~\cite{SPT-3G:2014dbx} and the 6-m Atacama Cosmology Telescope (ACT)~\cite{ACT:2020gnv} have similar science goals but in their recent observations have taken advantage of the natural conditions of their site.  The sky at the South Pole never sets but instead rotates around the axis. SPT thus focuses on observing small, always-visible patches deeply. ACT is at mid-latitude (23$^\circ$ S), so it can access large areas of the sky. ACT thus focuses on observing large areas of the sky to shallower depths. These optimized strategies enable SPT to measure the CMB temperature, polarization, and lensing modes to lower noise levels and find smaller galaxy clusters and fainter point sources. Meanwhile, ACT can measure more CMB and lensing modes, reducing the cosmic variance on the measurements, find a larger number of large and bright but rare clusters and sources, and have significant overlap with a number of optical surveys.

\begin{figure}[t]
    \centering
    \includegraphics[width=\columnwidth]{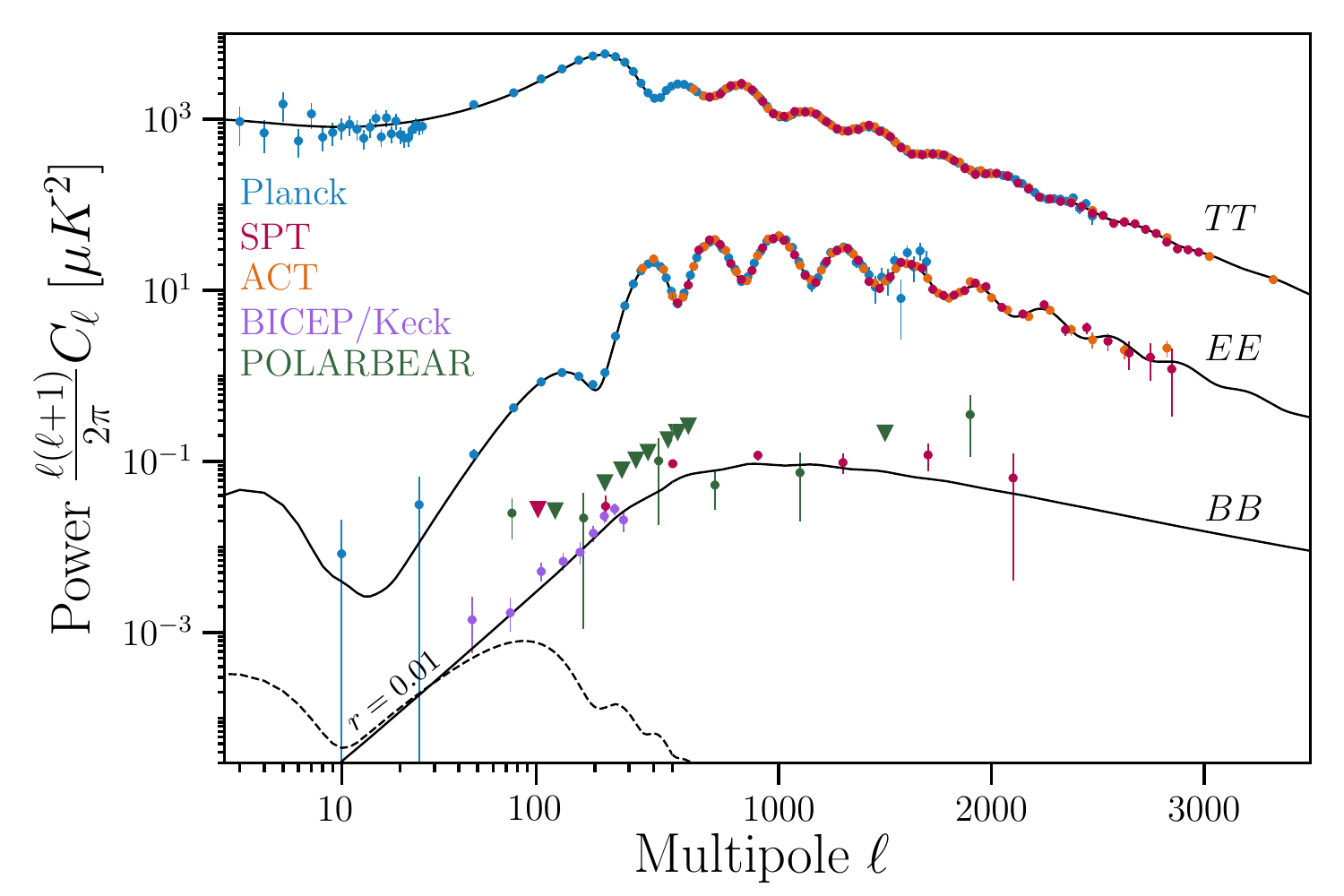}
    \caption{Recent measurements of the power spectra from Planck and selected ``Stage 3'' ground-based experiments \citep{planck18_spectra,dutcher21,story13,Choi:2020ccd,bk18,PB17,POLARBEAR:2022dxa}. The solid black lines show the CMB angular power spectra for the best-fit $\Lambda$CDM model while the black dashed line shows the primordial gravitational wave contribution corresponding to a tensor-to-scalar ratio parameter of $r=0.01$.}
    \label{fig:CMB_spectra_current}
\end{figure}

Figure~\ref{fig:CMB_spectra_current} shows the recent measurements from the ``Stage 3'' experiments, and how well they are fit by the temperature, $E$-mode, and $B$-mode lensing spectra of a six-parameter $\Lambda$CDM model.

\subsection{South Pole Observatory and Simons Observatory}
Current experiments are sited at the best observing locations on Earth, and their detectors are photon-noise limited. Thus improved measurements cannot be achieved through better detector sensitivity, but with more detectors. The need for %trend towards 
increasing detector counts has led to collaboration and consolidation among the experimental groups since detectors drive the costs of new instruments. (Fig.~\ref{fig:CMB_stages}). %(Fig.~\ref{fig:facility_consolidation}). However, improved sensitivity alone will not be sufficient for these experiments. 
%At the sensitivity to $B$-mode measurements improves, 
In the search for $B$-mode polarization from primordial inflation, at the sensitivities afforded by ``Stage 3'' experiments, it also becomes important to characterize and remove (or ``delens") $B$-modes generated from $E$-mode polarization via gravitational lensing by large scale structure \cite{2022ApJ...926...54A}. The need for delensing motivates coordinated observing between small and large-aperture telescopes. %has become the norm.

By combining the %$B$-mode sensitivity of the 
well-controlled systematics of small-aperture telescopes with the delensing capabilities of the large-aperture telescopes, the South Pole Observatory and Simons Observatory~\cite{SimonsObs} use the strengths of each telescope type to improve measurements of the tensor-to-scalar ratio.  In each project, detectors will number $\sim$50k-60k, an increase over the $\sim$10k in the individual ``Stage 3'' projects. The South Pole Observatory will combine measurements from the BICEP Array ~\cite{biceparray2020} and SPT-3G experiments ~\cite{sobrin21} operating at the South Pole. The Simons Observatory will field newly designed and constructed small and large-aperture instruments at Cerro Toco in the Atacama Desert, the same site as current and previous CMB experiments in the Atacama ~\cite{SimonsObs}.

\section{CMB-S4}
\label{sec:cmb-s4}

%\kmh{mostly from the DSR exec summary.  Needs some revising.}

CMB-S4 is a ``Stage 4'' CMB experiment with a rich and diverse set of scientific goals across four major themes \cite{2019arXiv190704473A,2016arXiv161002743A}: primordial gravitational waves and inflation, the dark Universe,  mapping matter in the cosmos, and the time-variable millimeter-wave sky. The CMB-S4 project aims to cross science-driven thresholds for inflation science and light relics. These goals are captured by two of the Project's top-level science requirements:
\begin{itemize}
\item CMB-S4 shall test models of inflation by putting an upper limit on $r$ of $r \le 0.001$ at 95\% confidence if $r = 0$, or by measuring $r$ at a $5\sigma$ level if $r>0.003$.  (Fig.~\ref{fig:nsrp01}.)

\item CMB-S4 shall determine $N_{\rm eff}$ with an uncertainty $\le 0.06$ at the 95\% confidence level.
\end{itemize}
Coupled with further science requirements on galaxy clusters and mm-wave transients, CMB-S4 will provide broad insights for high energy physics, cosmology and astrophysics.

%also have the capability to measure neutrino mass and gravitational lensing.
CMB-S4 will also measure of the sum of neutrino masses using multiple means, notably through measurements of gravitational lensing of the CMB and the abundance of SZ-detected galaxy clusters.
When combined with baryon acoustic oscillations from DESI~\cite{Font-Ribera:2013rwa}, and the current measurement of the optical depth to reionization from Planck~\cite{Planck2018}, CMB-S4 measurements of the lensing power spectrum will provide a constraint on the sum of neutrino masses of $\sigma(\sum m_\nu) = 24~\mathrm{meV}$, %and this would improve to $\sigma(\sum m_\nu) = 14~\mathrm{meV}$ if a future CMB survey (such as a space-based or balloon-borne experiment discussed below) were to provide a cosmic variance limited measurement of the optical depth. 
and this would improve to $\sigma(\sum m_\nu) = 14~\mathrm{meV}$ with better measurements of the optical depth (either from space-based or balloon-borne experiments, or through measurements of the kSZ signal by CMB-S4~\cite{2018PhRvD.FerraroSmith}).  Measurements of cluster abundances with CMB-S4 will provide similar constraining power, and taking the two measurements together enables a robust cross-check on the measurement of neutrino mass.  Furthermore, cluster abundance measurements with CMB-S4 will allow for a simultaneous measurement of neutrino mass and the dark energy equation of state (see Figure~\ref{fig:DE_Mnu_biref}). In addition to measurements of neutrino mass and dark energy, CMB-S4 will constrain various properties of dark matter~\cite{Snowmass2021:DarkMatter}.

%\kmh{Also mention dark matter, DE, and birefringence}

The CMB-S4 science requirements drive the project design. CMB-S4 will 
%Continuing the trend toward multi-resolution observatories, CMB-S4~\cite{2019arXiv190704473A} is a project 
%designed to 
conduct a combination of ultra-deep and deep-wide surveys, employing both low-systematics and high-resolution telescope facilities. %to address the CMB science cases in fundamental physics and astrophysics.  
The CMB-S4 detector payload will contain over 500,000 detectors, providing an order of magnitude more sensitivity than the previous generation.  CMB-S4 will be the first CMB experiment to represent a very large, community-wide effort. The formal CMB-S4 collaboration was established in 2018 with the ratification of the by-laws and election of the various officers including the collaboration Executive Team and Governing Board. As of spring 2022 the collaboration has 320 members, 76 of whom hold positions within the organizational structure. These members represent 114 institutions in 19 countries on 6 continents, including 27 US states.

The CMB community's advocacy for a single comprehensive experiment was endorsed by the 2014 report of the Particle Physics Project Prioritization Panel (P5) {\it Building for Discovery} and the 2015 NAS/NRC report {\it A Strategic Vision for NSF Investments in Antarctic and Southern Ocean Research.}  Most recently, CMB-S4 was strongly recommended by the 2020~Decadal Survey report {\it Pathways to Discovery in Astronomy and Astrophysics for the 2020s.}\cite{2021pdaa.book......}    The Technical, Risk, and Cost Evaluation (TRACE) ranked the project risk as medium-low.  The Decadal Survey recommendation reads ``The National Science Foundation and the Department of Energy should jointly pursue the design and implementation of the next generation ground-based cosmic microwave background experiment (CMB-S4).''
 
The CMB-S4 project is designed as a unified, single project that integrates the complex organization, policies, procedures, and support from the Department of Energy/Office of Science (DOE/SC/OHEP) and the National Science Foundation (NSF/AST/PHY/OPP).  In August 2020, the DOE selected Lawrence Berkeley National Laboratory (LBNL) to carry out the DOE roles and responsibilities in developing and executing the project. DOE has awarded CD-0, Approval of Mission Need, and is currently funding conceptual design studies and research and development.  NSF awarded the University of Chicago an MSRI-1 grant to perform preliminary design work in preparation to become a candidate for a Major Research Equipment and Facilities Construction (MREFC) award, in addition to a Mid-Scale Innovations Program (MSIP) award for preliminary design of a Large Aperture Telescope (LAT).  The project has developed the scope and a single Work Breakdown Structure (WBS), Organization, and Risk Registry for the entire project.

%The CMB-S4 design uses proven, existing technologies, developed and demonstrated over the last decades by CMB experimental groups, and scales them up to unprecedented levels. The design addresses the considerable technical challenges presented by the required scale-up of the instrumentation and by the scope and complexity of the data analysis and interpretation. Features of the design and plan include: superconducting detector arrays with well-understood and robust material properties and processing techniques; high-throughput mm-wave telescopes and optics with unprecedented precision and rejection of systematic contamination; full internal characterization of astronomical foreground emission; large cosmological simulations and improved theoretical modeling; and computational methods for extracting minute correlations in massive, multi-frequency data sets that include noise and a host of known and unknown signals. 

The timely completion of the project construction and robust support of experimental operations are important for the successful pursuit of the science case.  Also vital is preparatory support to the build up the analysis effort, so that it is ready for the flood of data that will arrive when operations commence.

\section{CMB-HD}
\label{sec:cmb-hd}

\begin{figure}[!b]
\centering
\includegraphics[width=0.47\textwidth]{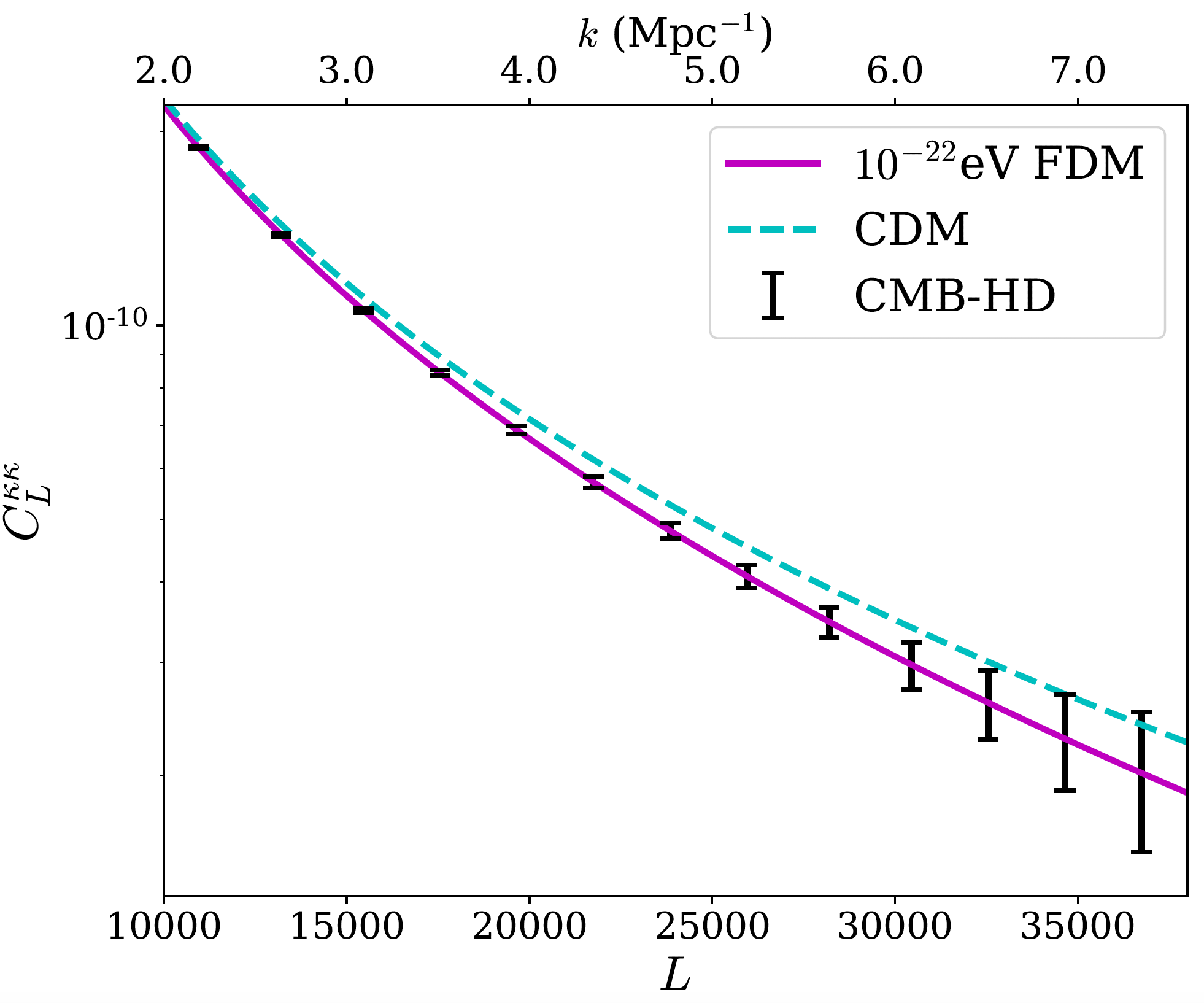}\hfill\includegraphics[width=0.5\textwidth,height=6.1cm]{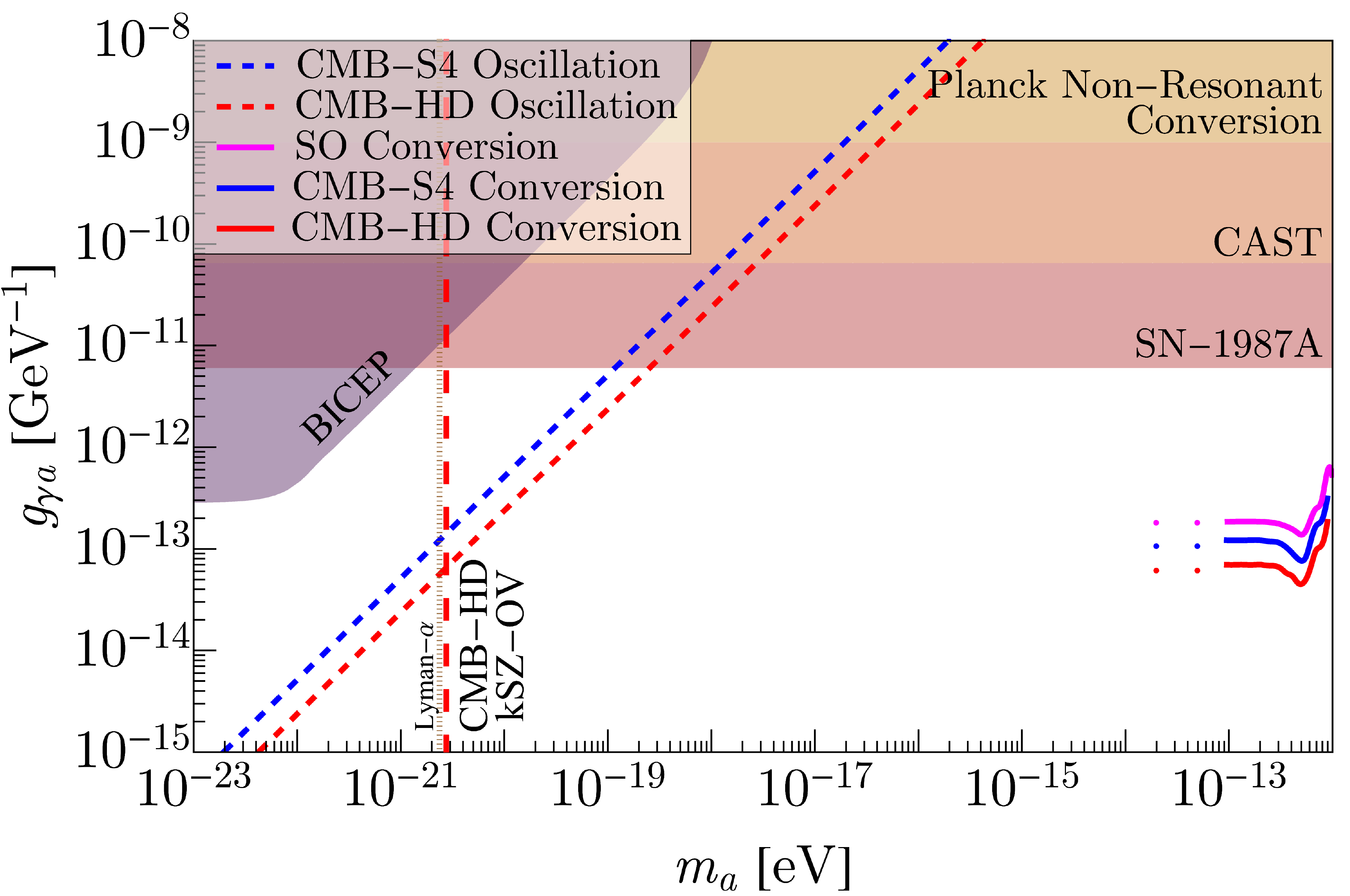}
\caption{{\it{Left:}} {\bf{Dark Matter:}} CMB-HD would generate via gravitational lensing a high-resolution map, out to $k\sim10~h$Mpc$^{-1}$, of the projected dark matter distribution over half the sky, and use that to probe the particle properties of dark matter.  {\it{Figure credit: \foreignlanguage{vietnamese}{Hồ Nam Nguyễn}}.} {\it{Right:}} {\bf{Axion-like Particles:}} Shown are forecasted constraints on axion-like particles (ALPs) from the resonant conversion of CMB photons into ALPs in the magnetic fields of galaxy clusters, from oscillation of the local CMB polarization angle, and from the kinetic SZ Ostriker-Vishniac (kSZ-OV) effect (the latter comparable to constraints from Lyman-$\alpha$ forest).  {\it{Figure credit: Suvodip Mukherjee, Michael Fedderke, Sayan Mandal, Gerrit Farren, and Daniel Grin.}}}
\label{fig:DMandAxions}  
\end{figure}

CMB-HD is a proposed future CMB experiment that would have three times the total number of detectors as CMB-S4 and $\sim 6$ times the resolution of current and planned high-resolution CMB telescopes~\cite{Abazajian:2019eic}, opening a new regime for mm-wave science.  CMB-HD would cross important thresholds for improving our understanding of fundamental physics, including the nature of dark matter, the light particle content of the Universe, the mechanism of inflation, and whether there is new physics in the early Universe beyond the Standard Model, as suggested by recent H0 measurements~\cite{Riess:2021jrx}. The combination of CMB-HD with contemporary ground and space-based experiments would also provide countless powerful synergies.

\begin{figure}[t]
\centering
\includegraphics[width=0.49\textwidth,height=6.1cm]{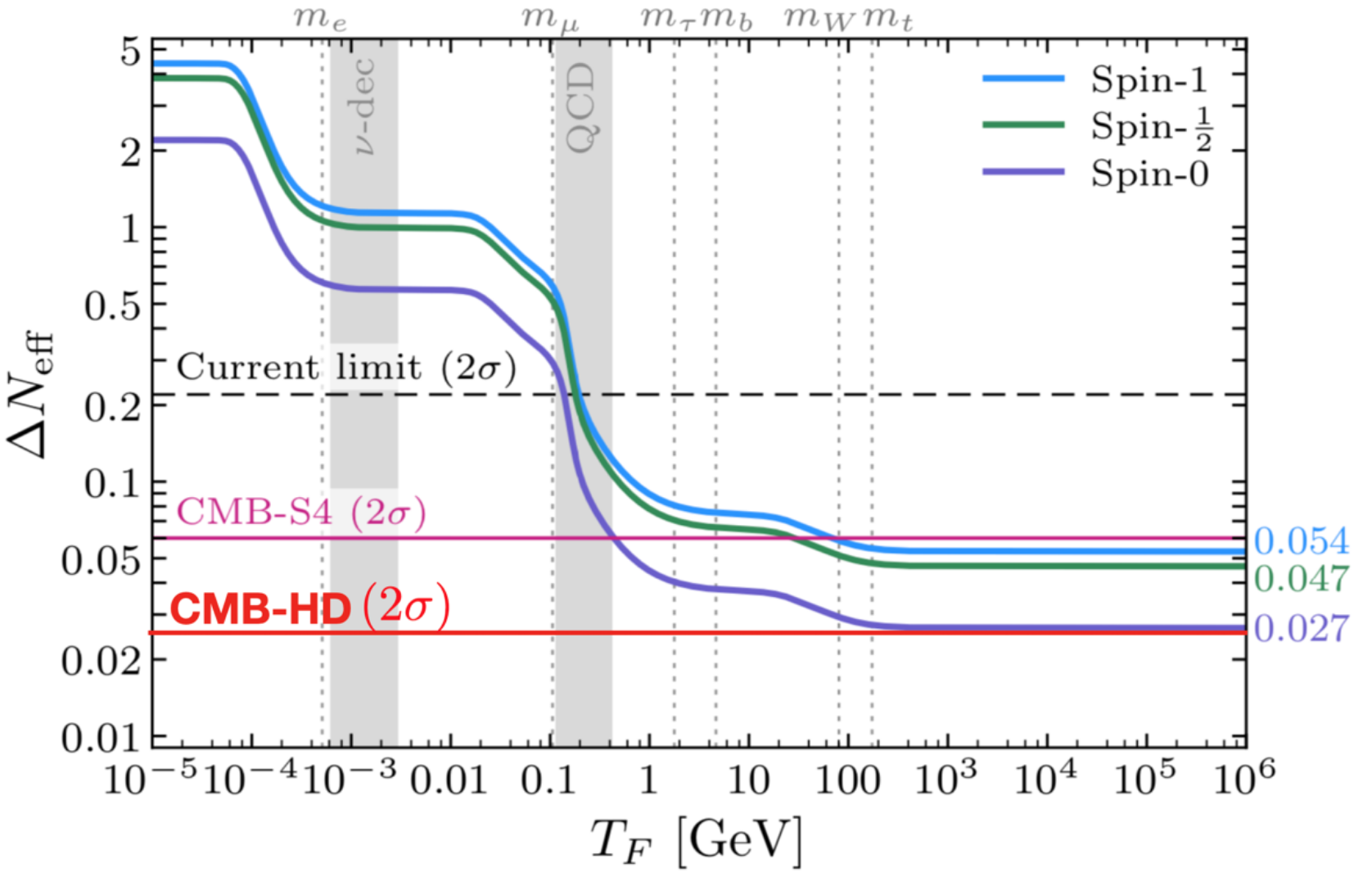}\hfill\includegraphics[width=0.47\textwidth]{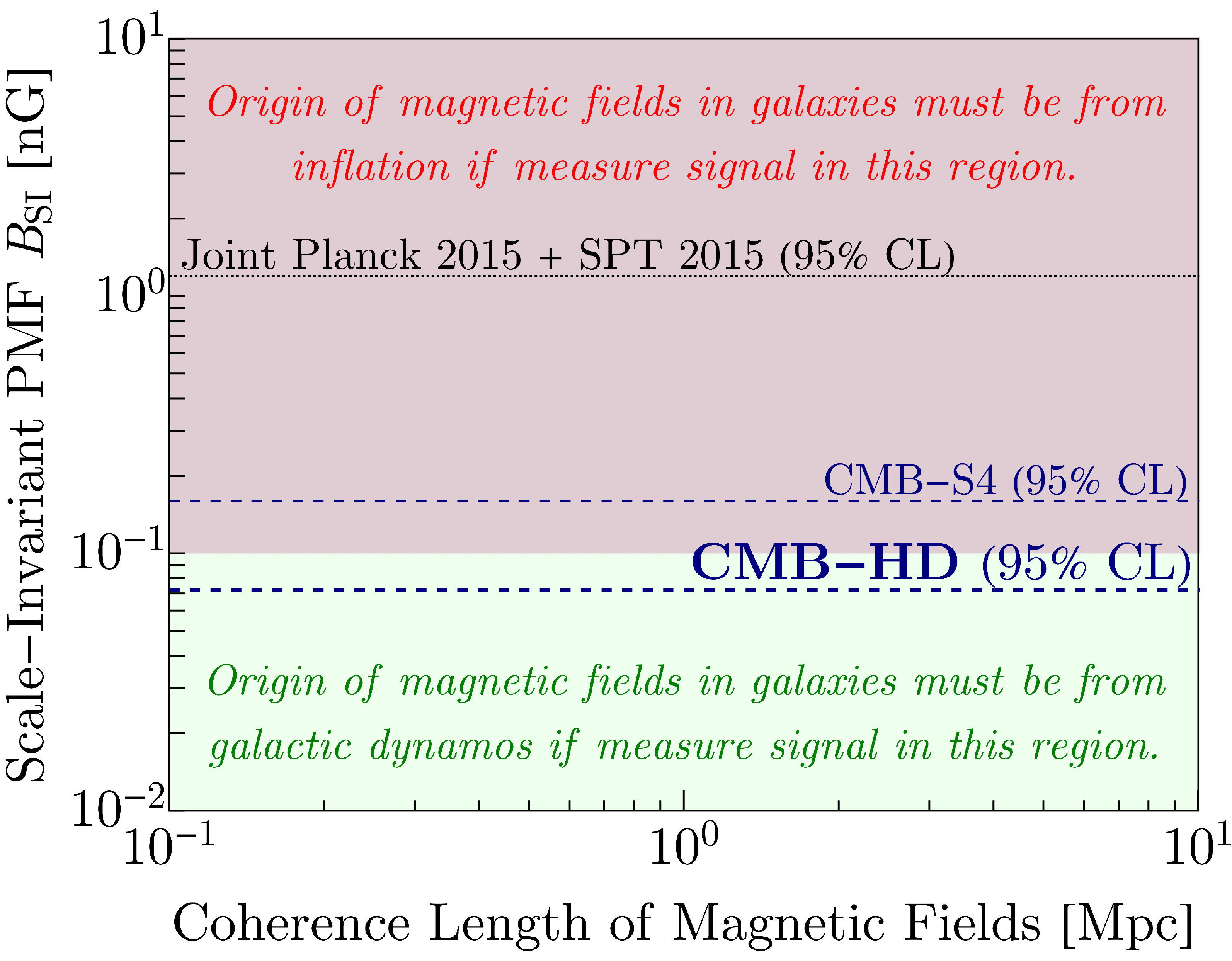}
\caption{{\it{Left:}} {\bf{Light Relic Particles:}} CMB-HD would achieve $\sigma({N_{\rm{eff}}}) = 0.014$, which would cross the critical threshold of 0.027.  {\it{Original figure from~\cite{Green:2019glg,Wallisch2018}; modified with addition of CMB-HD forecast.}} {\it{Right:}} {\bf{Inflation:}} CMB-HD could measure inflationary magnetic fields with an uncertainty of $\sigma(B)=0.036$~nG, crossing the critical threshold of $0.1\,\mathrm{nG}$ on Mpc scales~\cite{Mandal:2022tqu}.  {\it{Figure credit: Sayan Mandal.}}}
\label{fig:NeffandPMF}  
\end{figure}

\begin{figure}
\includegraphics[width=0.49\textwidth,keepaspectratio]{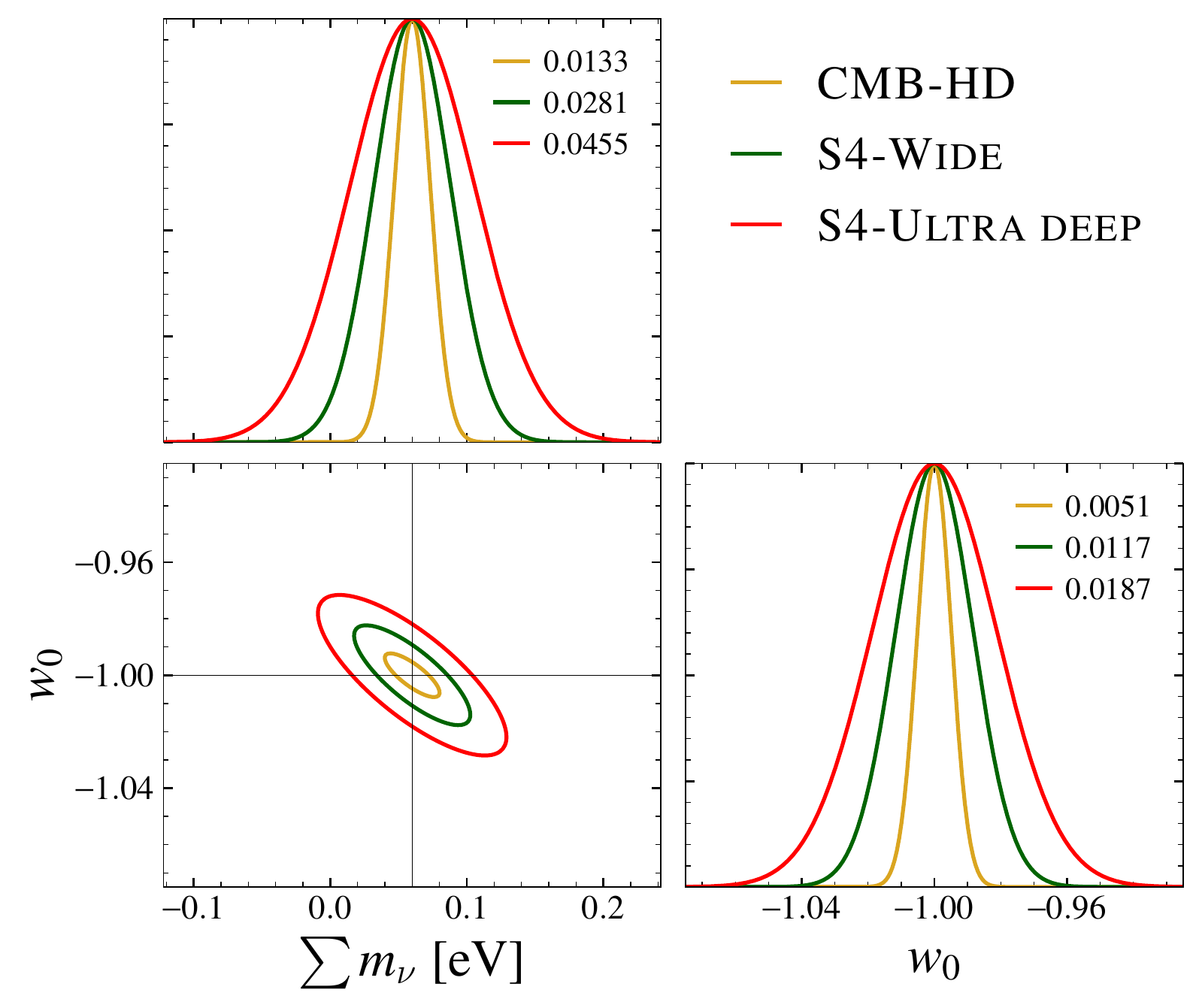}
\hspace{5mm}
\includegraphics[width=0.45\textwidth,height=7.0cm]{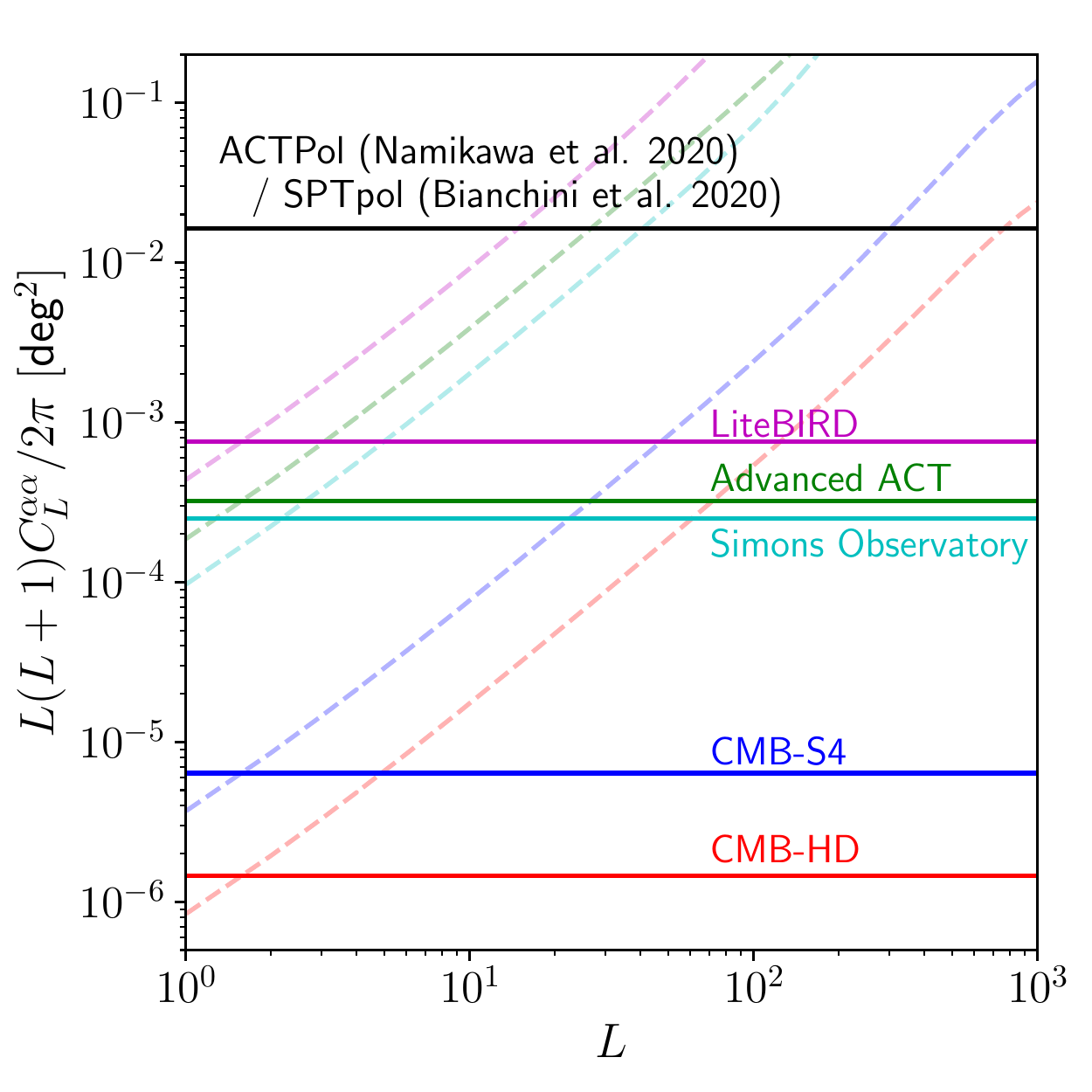}
\caption{{\it{Left:}} {\bf{Dark Energy and Neutrino Mass:}} Marginalized constraints on the dark energy equation of state parameter, $w_{0}$, and the sum of the neutrino masses, $\sum m_{\nu}$, combining primary CMB power spectra ($TT/EE/TE$) with cluster abundance measurements, and assuming a {\it Planck}-like $\tau_{\rm re}$ prior of $\sigma(\tau_{\rm re})= 0.007$\citep{raghunathan21a, raghunathan21b}. {\it{Figure credit: Srinivasan Raghunathan.}}
{\it{Right:}} {\bf{Beyond Standard Model:}} Forecasted $68\%$ CL bounds on anisotropic cosmic birefringence.  {\it{Figure credit: Toshiya Namikawa.}}}
\label{fig:DE_Mnu_biref}
\end{figure}

The concept for the CMB-HD instrument would consist of two new 30-meter-class off-axis crossed Dragone telescopes to be located on Cerro Toco in the Atacama Desert~\cite{Sehgal:2019ewc}.  Each telescope would host 800,000 detectors (200,000 pixels), for a total of 1.6 million detectors. The CMB-HD survey would cover half the sky over 7.5 years.  This would result in an ultra-deep, ultra-high-resolution millimeter-wave survey over half the sky with 0.5 $\mu$K-arcmin instrument noise in temperature (0.7 $\mu$K-arcmin in polarization) in combined 90 and 150 GHz channels and 15-arcsecond resolution at 150 GHz. CMB-HD would also observe at seven different frequencies between 30 and 350 GHz to mitigate foreground contamination. The key science targets of CMB-HD for fundamental physics are depicted in Figures~\ref{fig:DMandAxions},\ref{fig:NeffandPMF},\ref{fig:DE_Mnu_biref}.

Further details on the key science goals motivating the CMB-HD survey and the flowdown to measurement and instrument requirements are given in the Astro2020 Science White Paper~\cite{Sehgal:2019nmk}, Astro2020 CMB-HD APC~\cite{Sehgal:2019ewc}, Astro2020 CMB-HD RFI~\cite{Sehgal:2020yja}, and the Snowmass2021 CMB-HD White Paper~\cite{CMB-HD-Snowmass}. For further details see~\href{https://cmb-hd.org}{https://cmb-hd.org}.

\section{Space and Balloon Initiatives}
\label{sec:space}

As these ground-based efforts develop, it is important to provide the context of space-based and balloon-borne CMB efforts. %, even if these are outside the direct mandate of the Snowmass/P5 process.  
These instruments have synergies and complementarities to those on the ground \cite{2021pdaa.book......}.   Furthermore, the systematics that prove difficult for each type of experiment may be different, which makes cross-correlating them powerful.

On the ground, telescope apertures can be larger than can affordably be launched, providing higher resolution. Ground experiments may focus their sensitivity on a small portion of the sky for a long duration.  The ground provides instrument accessibility, a practical benefit that means the project lifetime is not limited by unreplenishable cryogens or fuel, and ground projects can be built in phases and easily upgraded.

In space, above the atmosphere, instruments have 5 to 10 times lower detector noise than on the ground, and they have access to the full frequency range of interest; they are not limited by atmospheric windows, nor by the atmospheric opacity and noise which are prohibitive above $\sim 300$~GHz.  
The excellent thermal stability at the L2 Lagrange point provides datasets that are more stable and less prone to thermal drifts, which results in stronger systematic error mitigation.  
Space missions can observe the entire sky, which is not possible from any single location on the Earth.  This gives them the ability to absolutely calibrate their data using the orbital-motion Doppler shift in the CMB-dipole. Thus previous space missions like WMAP and Planck have often been used to provide the crucial, overall calibration for ground based experiments. 
Finally, future all-sky, low-noise measurements of polarization from space/balloon will allow, in practice, for ground based instruments to calibrate directly on polarization, bypassing temperature altogether.  The absolute calibration of the space polarization data can be linked to the dipole at sub-percent levels through a number of independent avenues, both directly and through cosmological temperature--polarization cross correlations.

%and other ground and atmosphere induced effects dataset  

%On the ground, the detector counts can be higher, providing sensitivity, and the apertures can be larger than can affordably be launched, providing higher resolution.  The ground provides the practical benefit of instrument accessibility.  In space, there is no atmosphere, which has two implications.  First, the overall loading on the detectors is lower, so each detector is more sensitive.  Second, space missions are not limited to the narrow atmospheric windows that restrict ground observations.  This improves foreground removal and allows high-frequency observations.  Space missions can observe the entire sky, which is not possible from any single location on the Earth.  This gives them the ability to make an absolute calibration to the orbital-motion Doppler shift in the CMB-dipole.  Thus previous space mission like WMAP and Planck have often been used to provide the crucial, overall calibration for ground based experiments.

\subsection{LiteBIRD}
The Lite (Light) satellite for the study of $B$-mode polarization and Inflation from cosmic background Radiation Detection, or LiteBIRD, was selected by the Japan Aerospace Exploration Agency (JAXA) as a strategic large-class mission  \cite{2022arXiv220202773L}.  Following a late-2020s launch to the L2 Sun-Earth Lagrange point, LiteBIRD will map the sky for a three-year mission, covering the frequency range 34--448 GHz with 15 bands.  The resolution is 0.5$^\circ$ at 100 GHz.  With these capabilities, LiteBIRD can detect $B$-modes or alternatively set an upper limit of $r < 0.002$ at 95 percent confidence.

\subsection{NASA space and balloon missions}

\textit{Pathways to Discovery in Astronomy and Astrophysics for the 2020s} has endorsed only three Probe-scale missions; one of them is a future CMB Probe. Probes are \$1-billion-class missions. The report recommended preparatory work this decade with a possible implementation in the 2030s \cite[][]{2021pdaa.book......}.  
%
%The report quotes the subpanel on Electromagnetic observations from space saying ``Space observations will unquestionably be needed for the best foreground separation and lowest systematic errors on all angular scales,..."
% polarization to the limits set by the foregrounds would make a definitive statement about inflationary $B$-modes and other CMB fundamental physics science cases \cite[][]{2021pdaa.book......}. 
%The report also notes that to reach the limits promised by a future space mission preparatory work was needed in the following categories:
%\begin{itemize}
%    \item Improved detectors (filters and coupling, noise-limited detectors under space conditions, readout, cyrogenics).
%\item Better simulations of foregrounds and mitigation techniques.
%    \item Better simulations and mitigation of systematic errors.
%    \item Using knowledge of the above to optimize the flight system's design
%\end{itemize}
This work can lead to a competition for a Probe-class CMB mission design to be selected at the end of the decade.  A subpanel estimated that appropriate amount to spend over that decade on technology development is order \$100 million.

Groups have been considering potential designs for both imaging experiments, like PICO \cite{2019BAAS...51g.194H,2019arXiv190210541H}, and spectrometers to examine spectral distortions, like PIXIE \cite{2019BAAS...51g.113K}.

The PICO team was competitively selected by NASA in 2018 to produce a probe mission concept. The concept report \cite{2019arXiv190210541H}, which was endorsed by more than 200 scientists, presents an instrument with 1.4~m entrance aperture, two-mirror telescope coupled to a 0.1~K focal plane with 13,000 bolometric detectors spread over 21 frequency bands between 20 and 800 GHz. The resolution is between 1 and 38 arcminutes. Within the primary mission lifetime of 5-years PICO will conduct 10 redundant full sky surveys and give a required depth of 0.87~${\mu}$K arcmin (the estimated actual depth is 0.61~$\mu$K arcmin). Being equivalent to between 3300 and 6400 Planck missions, PICO will place 95\% confidence limit of $r<0.0002$ and provide a $5\sigma$ detection for $r=0.0005$, lowering $r$ limits by $\sim 5$ times compared to other existing or proposed projects. It is the only instrument that can measure the optical depth to reionization, $N_{\rm eff}$, and the sum of neutrino masses with precision not surpassed by any foreseeable instrument, all within the same single data set. The mission's deep polarization maps between 20 and 800 GHz will be used in many other astrophysical studies including for Galactic, Sunyaev-Zel'dovich, and Cosmic infrared background science. 

PIXIE would measure the large-scale B-mode signal to limits $r < 0.003$
(95\% CL) while providing a definitive full-sky map of the polarized 
diffuse dust cirrus on two-degree angular scale.  Measurements of spectral distortions provide a
unique window to the early universe, characterizing the amplitude of
primordial density perturbations on physical scales 3 orders of magnitude 
beyond those accessible to CMB anisotropy.  Such measurements provide
important constraints on primordial black holes and open a broad window
of discovery for dark matter interactions.

NASA's investment in the development of space missions includes the support of the scientific ballooning program.  Sub-orbital missions supported through that program have played an important part in developing detector technologies, testing instruments in remote conditions and near-space environments, and in training of future scientists.  They have made some of the most successful measurements of the CMB anisotropy.  The scientific balloon platform represents a unique capability that enables measurements in two categories: (1) observations at frequencies above $\sim 220$ GHz and (2) measurements probing the largest angular scales.   These frequencies and angular scales are more difficult to access from the ground than from the stratosphere due to atmospheric contamination.   Past missions, including MAXIMA, Boomerang, Archeops, Arcade, MAXIPOL, and Spider have published measurements of the CMB temperature and polarization derived from data more similar to that of space-based missions than ground based telescopes. Future missions, such as PIPER and Taurus, promise to extend polarization measurements of both the CMB and Galactic foregrounds, to the largest angular scales and to frequencies well above the peak of the CMB spectrum.  These high frequencies are critical for characterizing Galactic dust emission.   The high instantaneous sensitivity of the detectors and negligible impact of atmospheric emission that characterize the balloon borne instruments result in data for which the dominant systematic effects are distinct from those of ground-based telescopes.

\subsection{European efforts}

The European Space Agency (ESA) organized a consultation of the scientific community to define its scientific priorities for 2035-2050.  This ``Voyage 2050'' process defined the science themes for the next three large-class (L-class, $\sim 1$ billion Euro) missions, but did not yet define the specific mission concepts.  A coordinated series of whitepapers proposed science themes on spectral distortions \cite{2021ExA....51.1515C}, on using the CMB as a backlight for mapping structures and their baryon content \cite{2021ExA....51.1555B}, on cosmology via probing structure evolution across cosmic times with line-intensity mapping \cite{2021ExA....51.1593S}, and on an L-class mission to 
%accomplish them all, with various compromises
target them all, with compromises to be made between the different science cases
\cite{2021ExA....51.1471D}.  The \textit{Final recommendations from the Voyage 2050 Senior Committee} \cite{Voyage2050} selected ``New Physical Probes of the Early Universe'' as one of the science themes, and recommended ``a Large mission deploying gravitational wave detectors or precision microwave spectrometers to explore the early Universe at large redshifts.''  The same report mentions both the ``backlight'' and line-intensity mapping science cases as potential medium-class missions (M-class, $\sim 0.5$ billion Euro), which perhaps could be implemented together in a single mission.

In December 2021, ESA issued a new call for an M-class mission for launch in 2037.  Groups submitted three CMB-related proposals along similar science themes to the Voyage 2050 process.  
\begin{itemize}
    \item LISZT, a mission to probe structures at $2.1 \leq z \leq 5.3$ with [C-{\sc ii}] line-intensity mapping and observe Sunyaev-Zel'dovich effect and dust polarization at frequencies $>300$ GHz, complementary to ground-based observations at lower frequencies. 
    \item FOSSIL, a mission to target CMB spectral distortions within $\Lambda$CDM and beyond \cite{2021ExA....51.1515C} with a design that is similar to PIXIE but without polarization sensitivity and at slightly enhanced sensitivity.
    \item A request for ESA partipation to LiteBIRD.
\end{itemize}
In the M-class competition, ESA will make phase-1 selections around April 2022.  The phase-2 deadline is July 15, 2022. ESA is moving quickly in an attempt to make final selections by the end of 2022.

Proposed European balloon missions, such as BISOU \cite{2021arXiv211100246M} and OLIMPO, plan to exploit the balloon platform to measure spectral distortions in the CMB and to measure the Sunyaev-Zeldovich effect.

%\section{Summary }
%\label{sec:summary}
%\kmh{The summary is coming soon.}

\bibliographystyle{unsrt}
\bibliography{snowmassCMB.bib}

\end{document}